\documentstyle[aps]{revtex}
\newcommand{\beq}{\begin{eqnarray}}
\newcommand{\eeq}{\end{eqnarray}}
\begin{document}
\title{{On the log correction to the black hole area law}}
\author{Amit Ghosh\footnote{amitg@theory.saha.ernet.in}
and
P. Mitra\footnote{mitra@theory.saha.ernet.in}\\
Saha Institute of Nuclear Physics\\
Block AF, Bidhannagar\\
Calcutta 700 064, INDIA}
\maketitle
\begin{abstract}
Various approaches to black hole entropy yield the area law 
with logarithmic corrections, many involving a coefficient 1/2,
and some involving 3/2.  It is pointed out here that the standard
quantum geometry formalism is not consistent with 3/2 and favours 1/2.

\end{abstract}
\bigskip
There has long been an association of the area of the horizon of a black
hole with an entropy \cite{Bek}. This was not
initially understood according to the Boltzmann definition of
entropy as a measure of the number of quantum states of a black
hole, because of the absence of a proper quantum theory of
gravity. As a first step, however, considering gravity to be a
statistical system, the na\"{\i}ve Lagrangian path integral was
seen quite early to lead to a partition function from which the
area law of entropy was obtained \cite{GH} in the leading
semiclassical approximation ignoring all quantum fluctuations.
Subsequent support was obtained from considerations of quantum
fields in black hole backgrounds \cite{'tHooft,Uglum}. The
entropy calculated for the fields may be regarded as an additional
contribution to the entropy of the black hole - matter system, and
the gravitational entropy of the black hole itself may be imagined
to get modified in this way. In these field theory calculations
the leading term has a divergent multiplicative factor with the
area of the horizon. This divergence
may be thought of as a contribution to the bare or classical
gravitational constant $G$, which is to be renormalized to a
finite $G_R$ in the presence of quantized matter fields.

Recently some statistical derivations of the area law have appeared in
more elaborate models of quantum gravity -- in string theory \cite{strom}
as well as in quantum geometry \cite{ash}. Even
though a complete and universally accepted quantum theory of gravity is not
quite at hand, both of these approaches can accommodate the
expected number of quantum micro-states of a black hole.

With the area law so well established for the entropy of
large black holes, it is not surprising that even
corrections to the area formula have been studied.
The area of the horizon of an extremal dilatonic black hole
vanishes, and in this case the matter field approach was seen to
lead to a logarithm of the mass of the black hole \cite{GM} in the
expression for the entropy. For black holes with non-vanishing
area, the logarithm of the area appears as a sub-leading term
after the dominant term proportional to the area. The
coefficient of the logarithm depends on the black hole and is
1/90 in the Schwarzschild case.
These coefficients are expected to be renormalized, as indicated above.
Logarithmic corrections to the {\it gravitational} entropy, with
coefficients which are negative,
appeared later in many models. One approach \cite{romesh} was related to
the quantum geometry formulation but eventually mapped the counting
problem to conformal blocks, leading to a 
negative coefficient of magnitude 3/2.
Another \cite{carlip} started from ideas about conformal
symmetry in the near-horizon degrees of freedom 
and considered corrections to the Cardy formula,
reaching the same coefficient. There were
variations on these themes \cite{gour,romesh2}.

On the other hand, there has been a conflicting set of calculations leading to
a negative coefficient with the smaller magnitude 1/2. Among these,
\cite{jing} has followed the same conformal symmetry approach as \cite{carlip},
but has dropped the assumption made there that the central charge
is independent of the black hole area: it has in fact turned out
to be proportional to the area.
String theorists too \cite{sudipta,krause} have obtained the value 1/2, 
which has also appeared in applications of statistical
mechanics \cite{bhaduri}. In view of this disagreement,
it becomes necessary to examine the derivations of 3/2 more critically.
Surprisingly, there has been no direct calculation in
the quantum geometry approach \cite{ash},
in spite of the continuing progress in this field: the derivation 
\cite{romesh} used indirect conformal methods.
As the leading expression for the entropy has been calculated 
directly in the quantum geometry approach \cite{ash}, 
it is not difficult to look at the subleading contribution in the same manner.
A bound which is readily derived is consistent with the value 1/2
but not with the value 3/2. It is argued that 1/2 is in fact the actual value.

The calculation of black hole entropy from quantum geometry
is a simple counting problem described in detail in \cite{ash},
whose notation we more or less follow.
We consider a section of a
spherically symmetric {\it isolated} horizon. 
There are non-zero spins $J_a~(a=1,...,N)$ associated with punctures
on this sphere.
We work in units such that
$4\pi\gamma\ell_P^2=1$, where $\gamma$ is a `free' parameter 
(the Immirzi parameter) and $\ell_P$ is the Planck-length.
The spins are said to be {\it permissible} if the quantity
$|{\bf J}|\equiv2\sum_1^N[J_a(J_a+1)]^{1/2}$
lies in the range
\beq K-\epsilon\le\;|{\bf J}|\;\le K+\epsilon\label{perm},
\eeq
where $K$ is an integer representing the horizon area in
the above unit and $\epsilon\ll K$ 
compensates for the failure of $|\bf J|$ to be
an integer. Roughly, $|{\bf J}|\simeq K\sim N$, all
of these being large quantities.

For a permissible set of spins, the $a^{\rm th}$ puncture carries a
vector space of dimensionality $(2J_a+1)$,
so the net dimensionality of the representation is 
\beq d({\bf J})=\prod_a(2J_a+1) \;.
\eeq
There is a further restriction to be imposed: boundary conditions
require that
\beq\sum_a2m_a=0\quad {\rm mod}~ K
\label{m}\eeq
for each allowed configuration. So the physical `degeneracy' is
\beq d=\sum_{{\rm Permissible}\;\bf J}d_{\rm phys}({\bf J})\;,
\qquad{\rm where}\quad d_{\rm phys}({\bf J})=\int_{-\pi}^{\pi}
{d\theta\over 2\pi}\;\prod_a\sum_{m_a=-J_a}^{J_a}\exp(2im_a\theta)
\;.\label{deg}\eeq 
It may be noted here that $N<K$, so mod $K$ does 
not contribute to the counting. Since $d_{\rm phys}({\bf J})\le 
d({\bf J})$ for each permissible configuration, clearly $d$ obeys 
a bound 
\beq d\le\sum_{{\rm Permissible}\;\bf J}d({\bf J})\;.\label{upper}
\eeq

(\ref{upper}) has been used to put an upper bound on $d$ \cite{ash}:
\beq S=\ln d\le
{K\over\sqrt 3}\ln 2+O(K)\;,\quad\lim_{K\to\infty}{O(K)\over K}=0\;.
\eeq
However, we shall concentrate on a lower bound, which, as in \cite{ash},
can be obtained by considering
all spins to be $J_a=1/2$. Then, $|{\bf J}|=N
\sqrt 3$. Clearly, for $\epsilon\ge \sqrt 3$ it is always
possible to find an even $N$ obeying (\ref{perm}). 
The number of physical states or the degeneracy can be easily 
calculated from (\ref{deg}) $$d_{\rm phys}({\bf 1/2})=\left
(\matrix{N\cr N/2}\right)\;{\rm for\;even\;}N\;,\qquad 
d_{\rm phys}({\bf 1/2})=0\;{\rm for\;odd\;}N\;.$$
Thus, the entropy
\beq \ln d\;\ge \ln\left(\matrix{N\cr N/2}\right)\;,
\qquad N\in\;[{K\over\sqrt 3}-1,{K\over\sqrt 3}+1]\;.\label{lower}
\eeq
An estimate of the right hand side (\ref{lower}) can be made (cf \cite{ash})
with the Stirling approximation
\beq N!=N^N(2\pi
N)^{1/2}e^{-N}\left(1 +O({1\over N})+\cdots\right)\;.
\eeq
One obtains
\beq \ln d\ge\; N\ln 2-{1\over 2}\ln N+O(1)\;.
\label{result1}
\eeq
Now the two bounds on $N$ can be
exploited to obtain a bound on $\ln d$.
\beq N\ge
{K\over\sqrt 3}-1 &=\!\Rightarrow& N\ln 2\ge {K\over\sqrt 3}\ln 2
+O(1)\nonumber \\ N\le {K\over\sqrt 3}+1 &=\!\Rightarrow& -{1\over 2}\ln N
\ge -{1\over 2}\ln K+O(1/K)\;.
\eeq
Combining the two inequalities,
one gets
\beq S=\ln d\ge {K\over\sqrt 3}\ln 2-{1\over 2}\ln K+O(1)\;.\label{result2}
\eeq
(\ref{result2}) clearly shows, after
conversion of the integer $K$ into the area $A$ in appropriate units, that
there is a lower bound on the entropy of a spherically
symmetric isolated horizon:
\beq S\ge {A\over {\rm constant}} -{1\over 2}\ln A +O(1)\;.
\label{bound}\eeq

In the above derivation of the bound, only punctures with spin $J=1/2$ have been
considered, the reason being the dominance of spin $1/2$
in the number of physical micro-states 
\beq d=\sum_{\bf J}d_{\rm phys}({\bf J})\chi^\epsilon(K-|{\bf J}|)
\eeq 
where $\chi^\epsilon(x)$ is
the characteristic function for the interval $[-\epsilon,\epsilon]$.
This constraint (\ref{perm}) reveals that the number of punctures $N$ 
decreases as the spin $J_a$ at each puncture increases. An extreme case 
is one large $J$ at a single puncture. That reduces the degeneracy 
to one, $d_{\rm phys}=1$, since only the $m=0$ state contributes. 
However, 
the analysis of other intermediate configurations $\bf J$ is involved. 
We present now some estimates of the
contribution of other configurations to the entropy.  
 
If every puncture is associated with a common spin $j$, then 
$d_{\rm phys}({\bf j}\,)\le (2j+1)^{N_j}$ where $N_j\sim K/
[2\sqrt{j(j+1)}]$. So 
\beq \ln d_{\rm phys}({\bf j}\,)
\le K{\ln (2j+1)\over 2\sqrt{j(j+1)}}\;.
\eeq
If the constraint (\ref{m}) is implemented, one gets
\beq \ln d_{\rm phys}({\bf j}\,)
=  K{\ln (2j+1)\over 2\sqrt{j(j+1)}} -j\ln K +O(1)\;.  
\eeq
A logarithmic correction with the same coefficient was obtained
earlier in a different context \cite{saurya}.
But the dominant term falls off as $j$ increases: the factor 
\beq {\ln (2j+1)\over\sqrt{j(j+1)}}
\label{j} \eeq
has a maximum at $j= 1/2$ (if $j=0$, for which it is undefined,
is excluded). Consequently, 
the contribution of a configuration with $j>1/2$ to the entropy falls 
off at least like $\exp(-cK)$ where $c>0$, in comparison to a
configuration with $j=1/2$. E.g., $j=1$ 
produces a correction that falls off at least like $\exp[-0.02 K]$
compared to the dominant term and vanishes rapidly for large $K$.

For mixed configurations, the falloffs are {\em not} exponential.
It is not difficult to see that if a $1/2$ in the $\bf 1/2$
configuration is replaced by $j>1/2$, the decrease of the expression
(\ref{j}) implies that 
the contribution to the entropy is reduced. Different
contributions like this produce a $O(1)$ factor which does not
affect the dominant or logarithmic terms. We plan to
present soon \cite{gm} detailed evidence in favour of an {\it 
equality} for the entropy with 1/2 as the coefficient for the logarithmic
correction.

In conclusion, we have derived here a lower bound on the entropy of a black hole
strictly following the quantum geometry formalism \cite{ash}.
The popular value 1/2 for the logarithmic coefficient, which we
will support more fully elsewhere, is consistent
with this bound (\ref{bound}), but the older value 3/2 is not.
It is of interest to understand why this bound is inconsistent
with the value obtained in \cite{romesh}, which was also motivated
by quantum geometry, though calculated through conformal methods.
The condition (\ref{m}), which originates from boundary
conditions imposed on the isolated horizon \cite{ash}, requires only a
projection of the spins to add up to zero. In the approach of \cite{romesh},
weaker boundary conditions get imposed,
corresponding to an enhancement of the `gauge' symmetry
from U(1) to SU(2) and a decrease in
the number of `physical' states. This is best seen in the counting in the
second paper in \cite{romesh}, requiring not just a projection of the
total spin, but also its other components to vanish. This reduces the
number of states slightly, leaving the dominant term unaltered, but
changing the coefficient of the negative logarithm to 3/2 instead of 1/2.

\bigskip

We thank Abhay Ashtekar for many helpful suggestions on improving the
presentation.


\end{document}